\documentclass[twocolumn]{aastex61}
\usepackage{hhline}

\newcommand{\expon}[1]{e^{#1}}
\newcommand{\e}[1]{\times 10^{#1}}
\newcommand{\tableend}{\\}

\accepted{January 23, 2018}
\submitjournal{AJ}

\shorttitle{30 GHz Disk-Averaged Jovian Spectrum}
\shortauthors{Karim et al.}

\begin{document}

\title{A Wideband Self-Consistent Disk-Averaged Spectrum of Jupiter Near 30 GHz and Its Implications for NH$_{3}$ Saturation in the Upper Troposphere}

\correspondingauthor{Ramsey L. Karim}
\email{rkarim@astro.umd.edu}

\author[0000-0001-8844-5618]{Ramsey L. Karim}
\affiliation{Department of Astronomy,
University of Maryland,
College Park, MD 20742, USA}
\affiliation{Department of Astronomy,
University of California,
501 Campbell Hall,
Berkeley, CA 94720, USA}

\author[0000-0003-3197-2294]{David DeBoer}
\affiliation{Department of Astronomy,
University of California,
501 Campbell Hall,
Berkeley, CA 94720, USA}

\author{Imke de Pater}
\affiliation{Department of Astronomy,
University of California,
501 Campbell Hall,
Berkeley, CA 94720, USA}

\author[0000-0002-3490-146X]{Garrett K. Keating}
\affiliation{Department of Astronomy,
University of California,
501 Campbell Hall,
Berkeley, CA 94720, USA}
\affiliation{Harvard-Smithsonian Center for Astrophysics,
60 Garden Street,
Cambridge, MA 02138, USA}

\begin{abstract}

We present a new set of measurements obtained with the Combined Array for Research in Millimeter-wave Astronomy (CARMA) of Jupiter's microwave thermal emission near the 1.3 cm ammonia (NH$_{3}$) absorption band. We use these observations to investigate the ammonia mole fraction in the upper troposphere, near $0.3<P<2$ bar, based on radiative transfer modeling.
We find that the ammonia mole fraction must be $\sim$2.4$\e{-4}$ below the NH$_{3}$ ice cloud, i.e., at $0.8 < P < 8$ bar, in agreement with results by \cite{2001Icar..149...66D, 2016Sci...352.1198D}. 
We find the NH$_{3}$ cloud-forming region between $0.3 < P < 0.8$ bar to be globally sub-saturated by 55\% on average, in accordance with the result in \cite{2005Icar..173..439G}.
Although our data are not very sensitive to the region above the cloud layer, we are able to set an upper limit of $2.4\e{-7}$ on the mole fraction here, a factor of $\sim$10 above the saturated vapor curve.
\end{abstract}


\section{Introduction} \label{s:introduction}
Microwave observations of Jupiter's atmosphere are generally dominated by pressure-broadened spectral features of ammonia gas in the troposphere \citep{1963Icar....2..228T}.
The most notable feature in this region of the gas giant's thermal spectrum is the 1.3 cm NH$_{3}$ inversion/absorption band, first reported as a single, broad line by \cite{1968ApJ...154.1077L} and found by \cite{1978Icar...35...44K} to be a ``diagnostic of the pressure and temperature profiles in the cloud-forming region of the Jovian atmosphere.''
As we continue to sample Jupiter's thermal spectrum around this band, we can better characterize the shape of this spectral feature, which, through proper analysis, provides us with a deeper understanding of the planet's vertical structure and improves Jupiter's utility as a radio calibrator.

Decades of disk-averaged, and more recently, spatially resolved, observations as well as {\sl in situ} probe measurements have contributed to our understanding of the planet's atmospheric structure,
initially revealing a relatively straightforward model involving an adiabatic atmosphere with roughly solar\footnote{We designate a solar model based on the most recent proto-solar elemental abundance estimates \citep{2009ARA&A..47..481A}. N/H2 is taken to be $1.48 \e{-4}$. Henceforth, the solar NH$_{3}$ abundance value (or volume mixing ratio) in Jupiter is $1.28 \e{-4}$.} NH$_{3}$ mole fraction (hereafter, ``abundance'') in the deep atmosphere, which is considered to be well-mixed.
The detailed radiative transfer analysis presented in \cite{1985Icar...62..143D} agrees, stating solar NH$_{3}$ abundance to within a factor of 2 until $P < 0.5$ bar, above which NH$_{3}$ drops by $\sim$10$^{3}$.
This depletion is consistent with NH$_{3}$ condensation following the saturated vapor curve.

Modern efforts to uncover more about the planet's vertical structure have also raised new questions \citep{2005Icar..173..425D}.
The Galileo probe mission of the 1990s presented {\sl in situ} measurements of the NH$_{3}$ abundance that arguably conflict with models derived from ground-based observations, creating what was described in that paper as the ``Galileo \- Ground-based Microwave Paradox.''
Probe measurements suggest that NH$_{3}$ abundance should be nearly 4$\times$ solar. 
Given the previously accepted model, this result was jarring to our understanding of the atmosphere's structure and dynamics.  To improve this understanding additional datasets are needed along with updated modeling using the latest measurements of the microwave properties of, primarily, ammonia gas.

This work presents new ground-based observations made over a very wide bandwidth with the Sunyaev-Zel'dovich Array (SZA), a subset of the CARMA interferometer in an array configuration that does not resolve the planet, so more accurate disk-averaged brightness temperatures may be determined.
It provides a single systematically-consistent dataset over its band (27--35 GHz) that may be used in conjunction with the many decades of observations of our largest planet ({\em e.g.} \citealt{1978Icar...35...44K}, \citealt{2003ApJS..148...39P}, \citealt{2011ApJS..192...19W}, \citealt{2005Icar..173..439G}, \citealt{2016Sci...352.1198D}).
Having a well-calibrated systematically-consistent dataset over this broad bandwidth complements the single absolutely calibrated point near 28.5 GHz in \cite{2005Icar..173..439G} (hereafter JG).
In addition to presenting their own absolutely calibrated datapoint from earlier work \citep{2003PhDT........98G}, JG corrects and discusses the 20--24 GHz observations from \citep{1978Icar...35...44K}, and the 22--94 GHz observations obtained with the Wilkinson Microwave Anisotropy Probe satellite (WMAP) (\citealt{2003ApJS..148...39P}; updated in \citealt{2011ApJS..192...19W}); they conclude that NH$_{3}$ is globally subsolar between $0.6 < P < 2$ bar and subsaturated by more than 50\% between $0.4 < P < 0.6$ bar.

Based on these measurements we fine-tune the model presented in \cite{2001Icar..149...66D} and extended in \cite{2016Sci...352.1198D}. Based on Galileo measurements, this model assumes the NH$_{3}$ abundance to be $4.5 \times$ solar ($5.72 \e{-4}$) in the deep (P $>$ 8 bar) atmosphere and then follows the saturated vapor curve within and above the cloud layers.
This model, described in more detail in Section \ref{s:model}, is henceforth referred to as the ``nominal'' model and serves as the base model for this work.

\section{Data} \label{s:data}
One of the last studies conducted with CARMA was a CO power spectrum survey which aimed to measure the CO(1-0) transition in redshift $\sim$3 galaxies (\citealt{2015ApJ...814..140K}; \citealt{2016ApJ...830...34K}, hereafter referred to as COPSS).
The compact, 8-element Sunyaev-Zel'dovich Array (SZA) subset of CARMA was used to take various field scans between 27--35 GHz to measure the redshifted CO(1-0) $\nu_{0} = 115$ GHz line.
COPSS used Mars as the primary calibrator, and Jupiter as the secondary calibrator.
It is these Jupiter observations, taken between December 2014 and the array's decommission in April 2015, that are used in this work.

The SZA is a set of eight 3.5-m elements sensitive to left circularly polarized radiation in a compact array \citep{2015ApJ...814..140K}.
Since it does not resolve the planet, the array is well-suited for accurate measurements of Jupiter's total flux density.
The calibration data from COPSS comprise flux densities spanning 15 channels from 27--35 GHz (0.8 to 1.1 cm) and 100 days between December 2014 and April 2015.
Thermal error estimates are typically of order $0.01$ Jy for a single day's worth of data on Jupiter.
Uncertainty in the absolute flux calibration is preliminarily estimated at $<5\%$ according to COPSS, but we present our own thorough investigation in Section \ref{s:error}.
CARMA in its entirety is a fairly spectrally stable instrument, so we expect relative uncertainty to be somewhat smaller than absolute uncertainty.

\subsection{Determination of Jupiter's Flux Densities}

Reduction of these data converts a time series of flux densities to a single brightness temperature on channel-by-channel basis.
The process also isolates and corrects for a variety of systematics throughout the process.
The result is 15 time- and disk-averaged measurements of Jupiter's brightness near the 1.3 cm ammonia absorption band.

\begin{figure*}
	\plotone{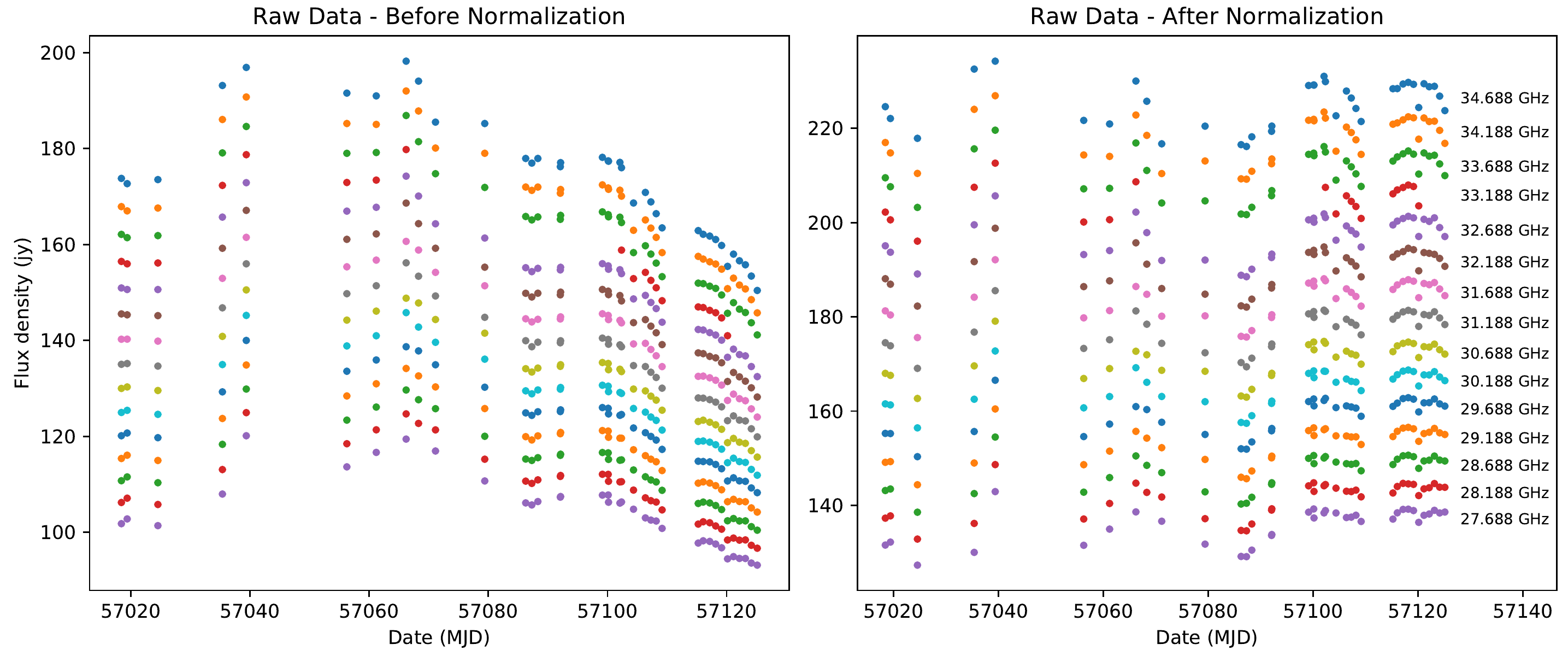}
	\caption{\label{fig:raw}Flux density measurements by time, before (left) and after (right) distance adjustment.\\ \\
    (Supplemental data for this figure are available in the online journal.)}
\end{figure*}

The effects of Jupiter's distance from Earth during the observational period are removed through distance normalization to the nominal value of 4.04 AU, ``flattening'' each channel's measurement as is made evident between the two panels in Figure \ref{fig:raw}.
This step facilitates examination of antenna gain error, indicated by the variance in the normalized points as well as the cross-channel behavior during individual observations.
Jackknife testing, discussed in more detail in Section \ref{s:jackknife}, reveals an additional artifact at this stage.
The series of flux densities for each channel show a positive linear correlation with Jupiter's elevation in the sky at the time of observation, indicating some potential issue with airmass calibration in the original measurements.
\begin{figure*}
	\plotone{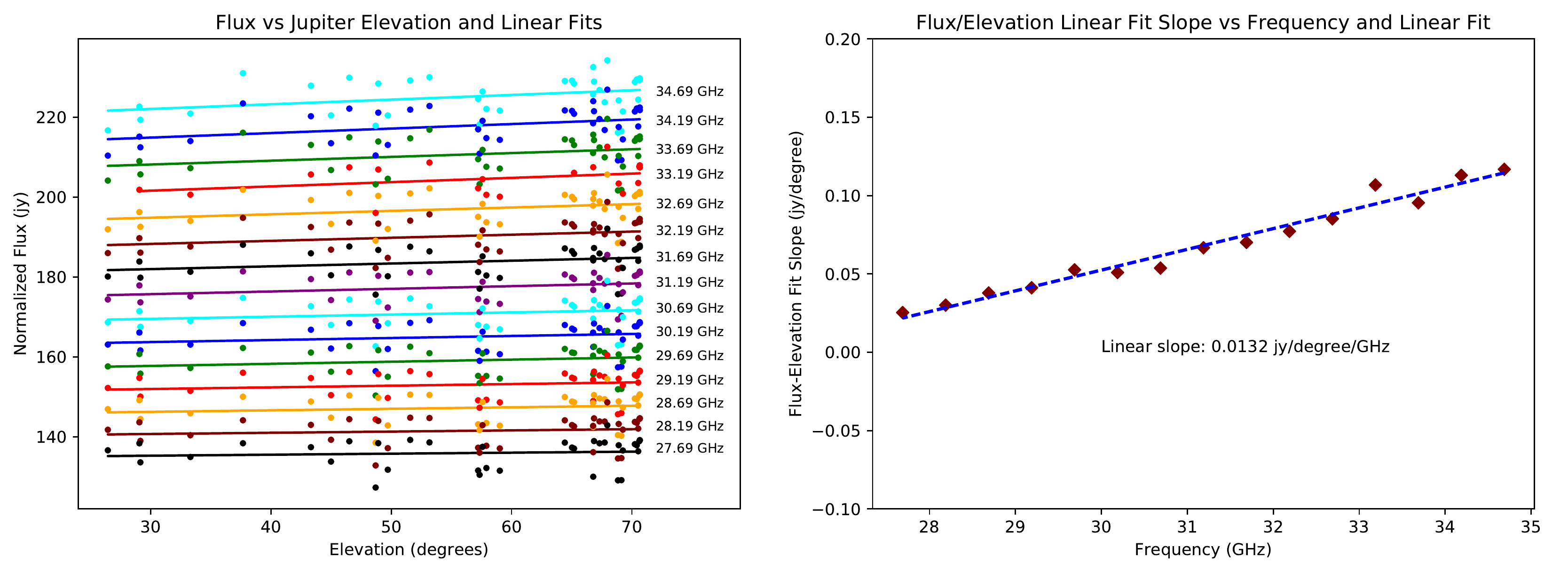}
	\caption{\label{fig:air}Indication of linear dependence of flux on elevation, and more importantly, linear dependence of flux-elevation slope on frequency. The left panel shows individual fits to each channel, sorted by Jupiter's elevation at observation time. The right panel shows these fits set against channel frequency and demonstrates the frequency dependence.}
\end{figure*}
The slopes of each channel's flux density are positive and linear with frequency, and therefore relatively easy to correct for, as demonstrated in Figure \ref{fig:air}; we assumed in this process that the measurements taken at higher elevations through thinner layers of atmosphere are more accurate.

The flux densities are averaged across time with weights corresponding inversely to the thermal error on each measurement.
Averages are calculated as
\begin{equation} \label{eq:fmeas}
F_{\nu,\,meas} = \langle F_{\nu} \rangle = \frac{ \sum_{i} w_{\nu, i} F_{\nu,i} }{ \sum_{i} w_{\nu, i} }
\end{equation}
where
\begin{equation} \label{eq:weight}
w_{\nu, i} = 1/\sigma_{\nu, i}^{2}
\end{equation}
such that $\sigma_{\nu, i}$ is the stated thermal error in Janskys on the $i$th measurement for a given channel.
Averaging the flux densities themselves is fairly straightforward; the rest of this section will discuss the application
of meaningful error estimates.

\subsubsection{Absolute ($\sigma_{A}$) and Relative ($\sigma_{R}$) Uncertainty} \label{s:error}

We define two distinct error measurements for the data set: absolute uncertainty $\sigma_{A}$ and relative uncertainty $\sigma_{R}$.
Absolute uncertainty quantifies our estimate of the overall calibration accuracy of the SZA, estimating how close the data are to the actual values.
This is determined by the systematics of the instrument.
An upper limit of $<5\%$ on this calibration error is quoted in COPSS, which incorporates results from \cite{2010ApJ...713...82S} in this estimate.

Relative uncertainty quantifies the stability of each channel with respect to the others, estimating internal consistency rather than absolute accuracy.
CARMA has strong spectral stability and can observe this feature in the strongly correlated cross-channel behavior displayed in Figures \ref{fig:raw} and \ref{fig:air}, which suggests that the $\sigma_{R}$ measurement, while allowing some independent uncertainty on points, is expected to be smaller than $\sigma_{A}$ and so will not have as great an effect on the overall position of the ensemble.
$\sigma_{R}$ is useful in defending the use of the relative structure of the ensemble as a comparably reliable tool to the absolute position as well as quantifying a model's match to this relative structure during the model-comparison process.
$\sigma_{A}$, by contrast, serves as a more conventional uncertainty estimate.

We approach the absolute uncertainty $\sigma_{A}$ estimate using statistical properties of the flux density ensemble along with their stated thermal errors.  The absolute uncertainty is the quadrature  sum of the weighted average of the individual data and the thermal noise of each measurement and may be written
\begin{equation}
\sigma_{A}^{2}(\hat{\sigma}, \hat{n}) = \hat{\sigma}^{2} + \hat{n}^{2}
\end{equation}
The first term is 
\begin{equation}
\hat{\sigma}^{2} = \frac{\sum_{i} w_{i} \big( F_{\nu,i} - \langle F_{\nu} \rangle \big)^{2} }{\sum_{i} w_{i}}
\end{equation}
The second term, the thermal contribution $\hat{n}$, is calculated as the reciprocal sum of thermal uncertainties for a given channel.
With $w_{i}$ defined as in Equation \ref{eq:weight}, $\hat{n}$ is
\begin{equation}
\frac{1}{\hat{n}^{2}} = \sum_{i} w_{i}
\end{equation}
In this way, we merge statistical 1-sigma error on a channel's flux ensemble with the thermal uncertainty of the points themselves and produce the value $\sigma_{A}$, our estimate of uncertainty on the absolute calibration of the instrument.
This we find to be closer to $\sim2\%$, which falls under the upper limit stated in COPSS.

Relative uncertainty $\sigma_{R}$ comprises both thermal contribution as well as the tendency of each channel to deviate from the others.
A perfectly stable instrument should exhibit a consistent cross-channel response to small, day-to-day variation in observed flux; unexpected behavior should be reflected in $\sigma_{R}$.
We compare observations across channels by normalizing each channel with its average\footnote{Until now, all averages have been across time, isolated to each channel.
In this section, that will no longer be the case, so averages will be marked as either channel averages ($\nu$) summed across all observations as before, or ``daily'' averages ($i$) calculated with sums across all channels, isolated to each observation.}.
The normalized measurements are denoted $f_{\nu, i}$.
\begin{equation}
f_{\nu, i} = \frac{F_{\nu, i}}{\langle F_{\nu, i} \rangle_{\nu}}
\end{equation}
This fraction is averaged across each observation to find the daily mean deviation from each channel's respective average.
A stable instrument's channel would exhibit minimal spread around this daily mean deviation, and any spread should be uncorrelated with channel.
We isolate the residuals $\delta_{\nu, i}$ from this daily average by subtracting the daily mean deviation from each observation set of normalized measurements, so as to exclude the daily deviation itself and examine each channel's deviation from this deviation.
These residuals are defractionalized by multiplying by the channel average.
\begin{equation}
\delta_{\nu, i} = (f_{\nu, i} - \langle f_{\nu, i} \rangle_{i}) \cdot \langle F_{\nu, i} \rangle_{\nu}
\end{equation}
Each channel's behavior is now quantified by its mean deviation from the daily average as well as the spread of these deviations, given by their variance across each channel.
During this examination, it was noted that the channels had a tendency to vary together in time; they often would pivot around the 31.188 GHz measurement, which remained stable to within $0.5$\% of the measurement value.
The frequencies at either extreme (27.688 GHz and 34.688 GHz) varied by no more than 6\%.
It is unclear what causes this linear variation, but it is worth mentioning that even though this estimate describes the uncertainty of these points relative to each other, it is an overestimate.
There remains some unidentifiable cross-channel dependency.

We introduce a thermal component as well, calculated by fractionalizing each thermal error by its accompanying measurement, averaging across each channel, and denormalizing with the channel average.
\begin{equation}
\sigma_{th} = \Bigg\langle \frac{\sigma_{i}}{F_{\nu,i}} \Bigg\rangle_{\nu} \langle F_{\nu, i} \rangle_{\nu}
\end{equation}

The final relative uncertainty measurement contains the thermal component $\sigma_{th}$ as well as the mean $\langle \delta_{\nu, i} \rangle_{\nu}$ and standard deviation $\Big\langle (\delta_{\nu, i} - \langle \delta_{\nu, i} \rangle_{\nu})^{2} \Big\rangle_{\nu}^{1/2}$ of the daily deviations by channel.
The third term, the variance of the channel's daily deviations, dominates the entire measurement and gives each channel a relative uncertainty on the order of a jansky.
\begin{equation}
\sigma_{R}^{2} = \sigma_{th}^{2} + \langle \delta_{\nu, i} \rangle_{\nu}^{2} + \Big\langle (\delta_{\nu, i} - \langle \delta_{\nu, i} \rangle_{\nu})^{2} \Big\rangle_{\nu}
\end{equation}

\subsubsection{Correction for the Synchrotron Radiation}

The relatively large SZA beam ($\theta_{B} \approx 11'$ full width at half max) contains flux from both thermal and non-thermal components.
In this frequency regime, synchrotron radiation dominates the non-thermal emission.
Since we are interested only in the thermal component, we must subtract out the synchrotron radiation from the total observed flux density.

Jupiter's dynamic synchrotron spectrum has been a subject of discussion since the 1970s when a series of observations suggested time variability in the low-frequency\footnote{\cite{1976JGR....81.3380K} uses 11--13 cm and 21 cm, considerably longer wavelengths than our 1 cm.} spectrum \citep{1976JGR....81.3380K}.
A survey of that spectrum from 74 MHz up to 8 GHz, described by \cite{2003Icar..163..434D}, suggests that the synchrotron contribution to the planet's radio spectrum drops off above 2 GHz, leading us to believe that, with frequencies around 30 GHz, our measurements contain negligible synchrotron-induced variability.
\cite{1976JGR....81.3380K} observes that fluctuations on the order of days did not exceed 10\% and explores variability on timescales of several years using 1--3 month averages, which implies that our 5-month average should capture an approximately constant period of synchrotron activity.
Under this assumption, we use a simplified and time-independent model to determine the contribution of synchrotron radiation on Jupiter's thermal spectrum.
The correction is purely arithmetic and thus does not propagate into uncertainties.

JG uses a value of 1.5 Jy for the synchrotron contribution to a 28.5 GHz measurement of the thermal spectrum, which has a value of about 145 Jy, based on work done by \cite{2003Icar..163..449D}.
In order to adjust extant data in the same frequency regime, JG adopts a relationship of $F_{\nu,\,synch} \sim \nu^{-0.4}$, leading to the local model
\begin{equation} \label{eq:synch}
F_{\nu,\,synch} =  (1.5 \ Jy)\Bigg(\frac{\nu}{28.5 \ GHz}\Bigg)^{-0.4}
\end{equation}
which we apply across our small frequency domain.

The synchrotron model (\ref{eq:synch}) is subtracted from the time-averaged flux density at each channel (\ref{eq:fmeas}), yielding thermal-only flux density measurements:
\begin{equation}
F_{\nu,\,thermal} = F_{\nu,\,meas} - F_{\nu,\,synch}
\end{equation}

\subsection{Conversion to Brightness Temperature and Cosmic Microwave Background (CMB) Adjustment}

The thermal radiation flux density, $F_{\nu,\,thermal}$, is converted to brightness temperature, $T_b$, via the Planck function.
The resulting $T_{b,\,meas}$ from a direct conversion is not yet indicative of the true temperature of the emitter---it is the contrast between the emitter and the microwave background.
Correction for this is made during conversion, following a similar adjustment made by \cite{2014Icar..237..211D}.
Observed thermal flux density is set equal to a combination of thermal brightness temperature and CMB contribution by the Planck function, as below, allowing $T_{cmb} = 2.725$ K.
\begin{equation}
F_{\nu} = \frac{2h\nu^3}{c^2}\Bigg( \frac{1}{\expon{h\nu/kT_b} - 1} - \frac{1}{\expon{h\nu/kT_{cmb}} - 1} \Bigg)
\frac{\pi R_{eq} R_{p}'}{D^2}
\end{equation}
where apparent polar radius is given by
\begin{equation}
R_{p}' = \sqrt{R_{eq}^{2}\sin^{2}{\phi} + R_{p}^{2}\cos^{2}{\phi}}
\end{equation}
Subearth latitude used here is $\phi = 0.15^{\circ}$, which is the average over a tight cluster of small, similar $\phi$ values over the four-month observation interval according to the JPL Horizons interface.

Working values at each major step as well as final measurements and associated error estimates are laid out in Table \ref{tab:1}.
The $T_b$ values, along with some combination of the $\sigma_{A}$ and $\sigma_{R}$ uncertainties, are appropriate for reproduction in future work.

\begin{table*}
\centering
\begin{tabular}{cc || c | c | c | c | c }
$f$ (GHz) & $\lambda$ (cm) & $F_{\nu,\,meas}$ (jy) & $F_{\nu,\,thermal}$ (jy) & $T_b$ (K) & $\sigma_{A}$ (K) & $\sigma_{R}$ (K) \\\hhline{==#=|=|=|=|=}
34.688 & 0.864 & 226.612 & 225.225 & 151.013 & 3.180 & 1.359 \tableend
34.188 & 0.877 & 219.250 & 217.855 & 150.385 & 3.143 & 1.186 \tableend
33.688 & 0.890 & 211.832 & 210.429 & 149.615 & 3.222 & 1.037 \tableend
33.188 & 0.903 & 206.006 & 204.594 & 149.876 & 2.531 & 1.244 \tableend
32.688 & 0.917 & 198.089 & 196.669 & 148.534 & 3.244 & 0.620 \tableend
32.188 & 0.931 & 191.047 & 189.618 & 147.707 & 3.228 & 0.510 \tableend
31.688 & 0.946 & 184.615 & 183.177 & 147.235 & 3.240 & 0.316 \tableend
31.188 & 0.961 & 178.155 & 176.708 & 146.635 & 3.293 & 0.146 \tableend
30.688 & 0.977 & 171.458 & 170.002 & 145.720 & 3.266 & 0.378 \tableend
30.188 & 0.993 & 165.546 & 164.080 & 145.347 & 3.382 & 0.395 \tableend
29.688 & 1.010 & 159.553 & 158.077 & 144.794 & 3.515 & 0.675 \tableend
29.188 & 1.027 & 153.335 & 151.849 & 143.911 & 3.584 & 0.769 \tableend
28.688 & 1.045 & 147.477 & 145.981 & 143.225 & 3.679 & 0.960 \tableend
28.188 & 1.064 & 141.586 & 140.080 & 142.369 & 3.767 & 1.147 \tableend
27.688 & 1.083 & 136.081 & 134.563 & 141.757 & 3.875 & 1.368 \tableend
\end{tabular}
\caption{\label{tab:1}Values at each frequency through major correction steps, after averaging. $F_{\nu,\,meas}$ and $F_{\nu,\,thermal}$ are normalized to 4.04 AU. $T_b$ values include CMB correction. Error estimates are given for the final $T_b$ values only.}
\end{table*}

\subsection{Jackknife Testing} \label{s:jackknife}
In order to investigate whether identifiable systematics are present, we conducted a set of jackknife tests during which we run arbitrarily selected halves of the time-series data through the analysis pipeline and observe the average resulting change in brightness temperatures.
The data are halved both on meaningless criteria as well as by criteria with more systematic potential.
We applied arbitrary-parameter tests with the data split along odd vs.\ even index, first vs.\ second half, and first and last quarters vs.\ central half.
We applied the more meaningful test of sorting along Jupiter's horizontal elevation at the time of observation.

It was mentioned earlier that what may be an airmass calibration error in the raw data was detected by one of these jackknife tests, specifically one using Jupiter's elevation.
After this correction, all subsequent jackknife tests amount to nothing more than noise at less than 2\% variation from the full-range values, indicating that we have identified all major systematics about which we have information.
The consistency of our measurements with the Gibson and WMAP points corroborates this, or at least suggests that we all suffer from the same unknown systematics.

\section{Radiative Transfer Analysis} \label{s:model}

As briefly discussed in Section \ref{s:introduction}, the main source of radio opacity in Jupiter's atmosphere is ammonia gas.
The following subsections outline our methods to determine the NH$_{3}$ abundance in that part of the atmosphere to which our measurements are sensitive.

\subsection{Nominal NH$_{3}$ Profile}

Our calculations use the radiative transfer code most recently used by \cite{2016Sci...352.1198D}; this code is based upon an atmosphere in thermochemical equilibrium, as described in detail by \cite{2005Icar..173..425D}.
As in \cite{2016Sci...352.1198D}, we assume for our nominal model that all constituents (NH$_{3}$, H$_{2}$S, CH$_{4}$, and H$_{2}$O) are enhanced by a factor of 4.5 above solar in the deep atmosphere ($P > 8$ bar), as observed by the Galileo probe for NH$_{3}$, H$_{2}$S, and CH$_{4}$ (\citealt{1998JGR...10322847F}, \citealt{1999BAAS...31.1154M}, \citealt{1998JGR...10322929S}, \citealt{2004Icar..171..153W}).
At higher altitudes, NH$_{3}$ will be partially dissolved in the water cloud ($\sim$7.3 bar), will form the NH$_{4}$SH cloud ($\sim$2.5 bar, \citealt{1990AdSpR..10...79D}), and at $P < \, \sim$0.8 bar will condense into its own ice cloud and follow the saturated vapor curve.
In our nominal model we thus assume an NH$_{3}$ abundance of $5.72 \e{-4}$ in Jupiter's deep atmosphere, which is diminished at altitudes at which clouds form.
We adopt a 100\% humidity within and above the NH$_{3}$ ice cloud in our nominal model.
This results in a constant abundance of $1.20\e{-7}$ near and above the tropopause. This NH$_{3}$ profile is shown by the blue curve in Figure \ref{fig:tp}b, and the resulting spectrum is shown by the blue curves in Figures \ref{fig:emission_wl} and \ref{fig:emission_freq}.

Note from the weighting functions in Figure \ref{fig:tp}a that our data are sensitive to $P < \, \sim$3 bar.
While our results should, strictly speaking, only apply down to a ``deep atmosphere'' cutoff defined by this sensitivity, we extend our results down to $P < 8$ bar in order to remain consistent with \cite{2016Sci...352.1198D} under the assumption that NH$_{3}$ abundance should remain roughly constant between $3 < P < 8$ bar---it should only increase (with increasing pressure) at the clouds described above.
Below $P > 8$ bar, we adopt the values as measured by the Galileo spacecraft \citep{2004Icar..171..153W}

\subsection{Model Generation Through Perturbation}

Starting with the nominal NH$_{3}$ abundance profile described above, we apply to it small, unity-order adjustments, which in turn gives us the ability to generate a range of theoretical spectra.
These spectra are compared to the available measurements in order to isolate a NH$_{3}$ abundance profile in maximal agreement with observations.
The spectra are generated using the radiative transfer software, pyplanet, described by \cite{2014Icar..237..211D} for its use on Neptune's atmosphere and most recently updated with the NH$_{3}$ line profile from \cite{2016Icar..280..255B}.
The software ignores potential opacities from clouds.

\begin{figure*}
	\epsscale{1.18}
	\plotone{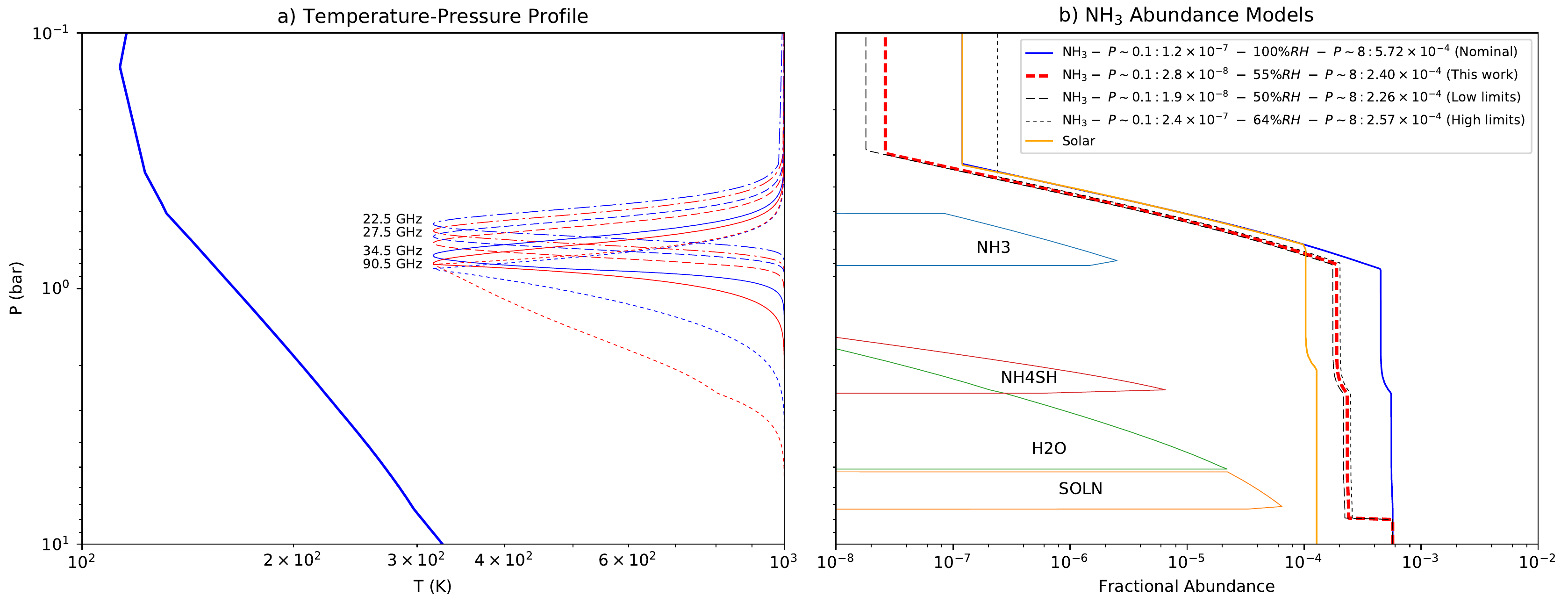}
	\caption{\label{fig:tp}a) The temperature-pressure profile (blue curve) used in our model of Jupiter's atmosphere. Several weighting functions are overlaid on the right edge of this figure. b) The fractional NH$_{3}$ abundances according to the solar (yellow) and nominal (blue) constituent abundance models plotted next to our best-fit model (red dashed line). The thinner black lines indicate global high- and low-abundance limits using the error bounds for each parameter in the 3-parameter grid fit. The lower bound to the NH$_{3}$ abundance near the tropopause is not certain; this plot uses the small value resulting from the Clausius-Clapeyron equation of state. Cloud levels are overlaid on the left hand side of this panel.}
\end{figure*}

We begin this process by separating the atmosphere into regions of altitude based on their radiative contribution to our measurements, according to the nominal NH$_{3}$ abundance model.
These contribution functions, shown in Figure \ref{fig:tp}a, are most prevalent between $0.5 < P < 0.8$ bar.
This layer coincides with the NH$_{3}$ ice cloud-forming region, over which altitude range NH$_{3}$ follows the saturated vapor curve, as indicated by the nominal abundance profile (blue curve in Figure \ref{fig:tp}b).
We apply to this region, defined by its plummeting NH$_{3}$ abundance, a constant humidity multiplier $RH$.
Through this parameter, we will tune humidity in the NH$_{3}$ cloud-forming region to fit observations.

We similarly treat the regions of constant NH$_{3}$ abundance above and below this layer, granting them their own modifying constants.
The sub-cloud region, defined as the region of approximately constant NH$_{3}$ abundance from the bottom of the NH$_{3}$ cloud down to $P \sim 8$ bar, is modified by the parameter $\alpha_{d}$.
As we consider our model not to apply below $P > 8$ bar, we jump the abundance below this point back to the nominal model, which is motivated by Galileo measurements sensitive to this deeper region.
\cite{2016Sci...352.1198D} applies this same practice.
The region of constant abundance above the tropopause is modified by the parameter $\alpha_{h}$.

These parameters, $RH$, $\alpha_{d}$, and $\alpha_{h}$, form a three-parameter grid across which we create slightly modified versions of the nominal abundance model.
Each element of the grid is run through the radiative transfer software to generate a spectrum, and each of these spectra is compared to the observations such that each point on the three-dimensional grid is associated with a $\chi^{2}$ value.
We search the parameter space, bounded by a physically reasonable range of unity-order constants, for a global minimum. This minimum is explored to within 0.5\% resolution of the nominal value.
The abundance profile associated with this minimum is taken to be the profile in maximal agreement with observational evidence. Uncertainty for each parameter is approximated as the region in which the local $\chi^{2}$ value is less than $2\times$ the minimum $\chi^{2}$ value.

\section{Results} \label{s:results}

Our measurements of Jupiter's atmospheric emission show a smoothly sloped curve on the short-wavelength side of the 1.3 cm NH$_{3}$ absorption band.
We compare our data with the surrounding WMAP and Klein \& Gulkis datasets, as adjusted in JG, as well as Gibson's original point at 28.5 GHz. The CARMA observations are consistent to well within 1\% with points from these existing measurements that fall between 27--35 GHz. In the 2--6 cm range we use VLA measurements from \cite{2016Sci...352.1198D} that have since been recalibrated \citep{2016AGUFM.P31D..08D}.

We implement the model-measurement comparison scheme discussed in the previous section and generate a grid of model spectra and corresponding $\chi^{2}$ comparison results.
The $\chi^{2}$ grid fitting we implement compares models only to the CARMA, WMAP, and JG measurements.
Comparisons to Klein \& Gulkis and VLA data are made after best-fit values have been obtained in order to verify the results.

\subsection{Sub-Cloud Abundance $\alpha_{d}$}

We examine the NH$_{3}$ abundance below the cloud-forming region relative to the nominal abundance value of $5.72\e{-4}$, and find a value of $2.40\e{-4}$ just above $P \sim 8$ bar, with uncertainty bounding it between $[2.26, 2.57] \e{-4}$.
Accounting for reductions at clouds, the $P \sim 0.8$ abundance is $1.89\e{-4}$.
Figure \ref{fig:emission_freq} demonstrates the individual contribution of $\alpha_{d}$ (dashed cyan line) to the final model (red line).
This contribution dominates far from the band center, where it fits especially well to the high frequency WMAP and lower frequency VLA points, both sensitive to the deeper atmosphere.

Considering that ours is a disk-averaged result, these measurements are consistent with results recently presented in \cite{2017GeoRL..44.5317L} using data from the Microwave Radiometer (MWR) experiment on the Juno spacecraft.
Latitudinally-resolved abundance measurements from the MWR, shown in \cite{2017GeoRL..44.5317L} Figures 3 and 4, at all latitudes except the more ammonia-rich Equatorial Zone tend towards this same $2$--$2.4\e{-4}$ from the NH$_{3}$ cloud at $P \sim 0.8$ bar down to about 10 bar, a regime consistent with our sub-cloud partition.

\subsection{Relative Humidity $RH$}

Humidity is examined relative to the saturation curve region in the nominal model, roughly between $0.3 < P < 0.8$ bar.
The nominal model follows 100\% humidity, so our results will be considered relative to a fully saturated model.
Best-fit results suggest a humidity of $56.5\%$, bounded between $[50.0\%, 63.5\%]$.
The dotted cyan line in Figure \ref{fig:emission_freq} shows the individual contribution of $RH$, which dominates closer to the band center and produces model spectra that are consistent with the lower frequency CARMA points and the Gibson point that are most sensitive to this pressure range.

JG states that the NH$_{3}$ abundance is, on average, sub-saturated by at least a factor of 2 at $P < 0.6$ bar.
Our results corroborate this  factor-of-2 sub-saturation  within our own pressure regime stated above, but add a tighter bound based on 4 independent sets of measurements.

\subsection{High Altitude Atmosphere Abundance $\alpha_{h}$}

The high altitude atmosphere abundance is examined relative to the nominal (saturated vapor curve) value of $1.2\e{-7}$.
Our model fits suggest a value of about $2.8\e{-8}$, nearly one fifth of the original, but yield an upper bound of $2.4\e{-7}$, twice the original value.
Since we are not very sensitive to these pressures, we are not able to provide a lower bound.

This result relies primarily on the lowest frequency WMAP measurement but has some input from the lower frequency CARMA measurements.
It is difficult to place any reasonable bound on the high altitude atmosphere abundance due to the lack of data near the band center; the Klein \& Gulkis measurements were deemed too uncertain for our comparison, but tend toward temperatures higher than the WMAP point.
Higher brightness temperatures at the band center would indicate a lower pressure departure from the saturation curve and consequently a smaller high altitude atmosphere abundance.
The precise number found in our analysis is likely meaningless -- nevertheless, it is quite possible that the high atmosphere abundance should be smaller than its value in the nominal model.

\begin{figure*}
	\plotone{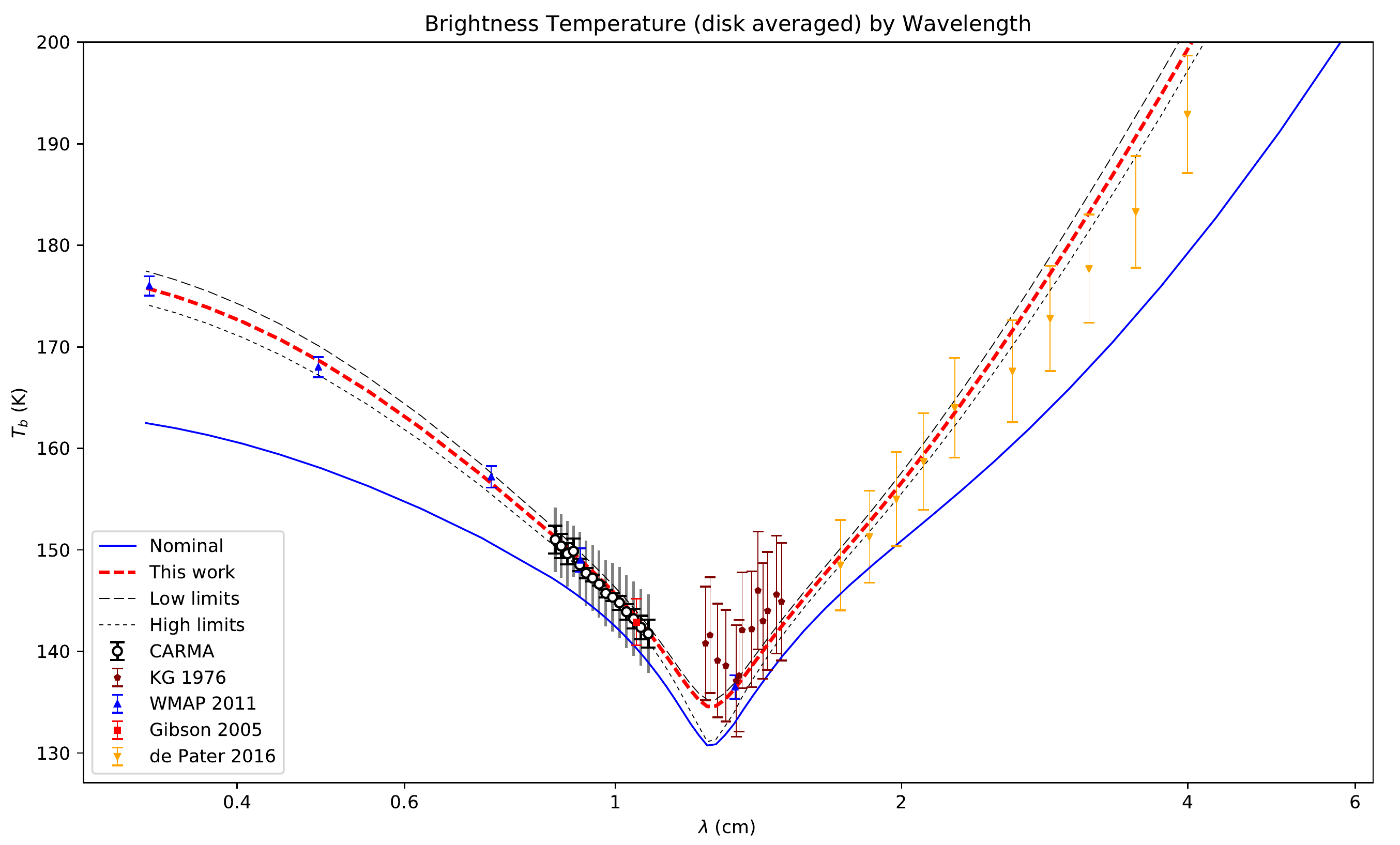}
	\caption{\label{fig:emission_wl}CARMA measurements of disk-averaged brightness temperature together with extant measurements (\citealt{1978Icar...35...44K}, \citealt{2011ApJS..192...19W}, \citealt{2005Icar..173..439G}, \citealt{2016AGUFM.P31D..08D}). Nominal model is included, as well as the model we suggest in this work, using parameters discussed in Section \ref{s:results}, and the global high- and low-abundance limits. The lower limit for high altitude atmospheric abundance is, as discussed in the text, not stated in this work. CARMA measurements are shown with a relative certainty at the capped black error bars and absolute uncertainty at the larger, uncapped grey bars. Note the uniformity in absolute certainty and the frequency-dependent variation in relative uncertainty.}
\end{figure*}

\begin{figure*}
	\plotone{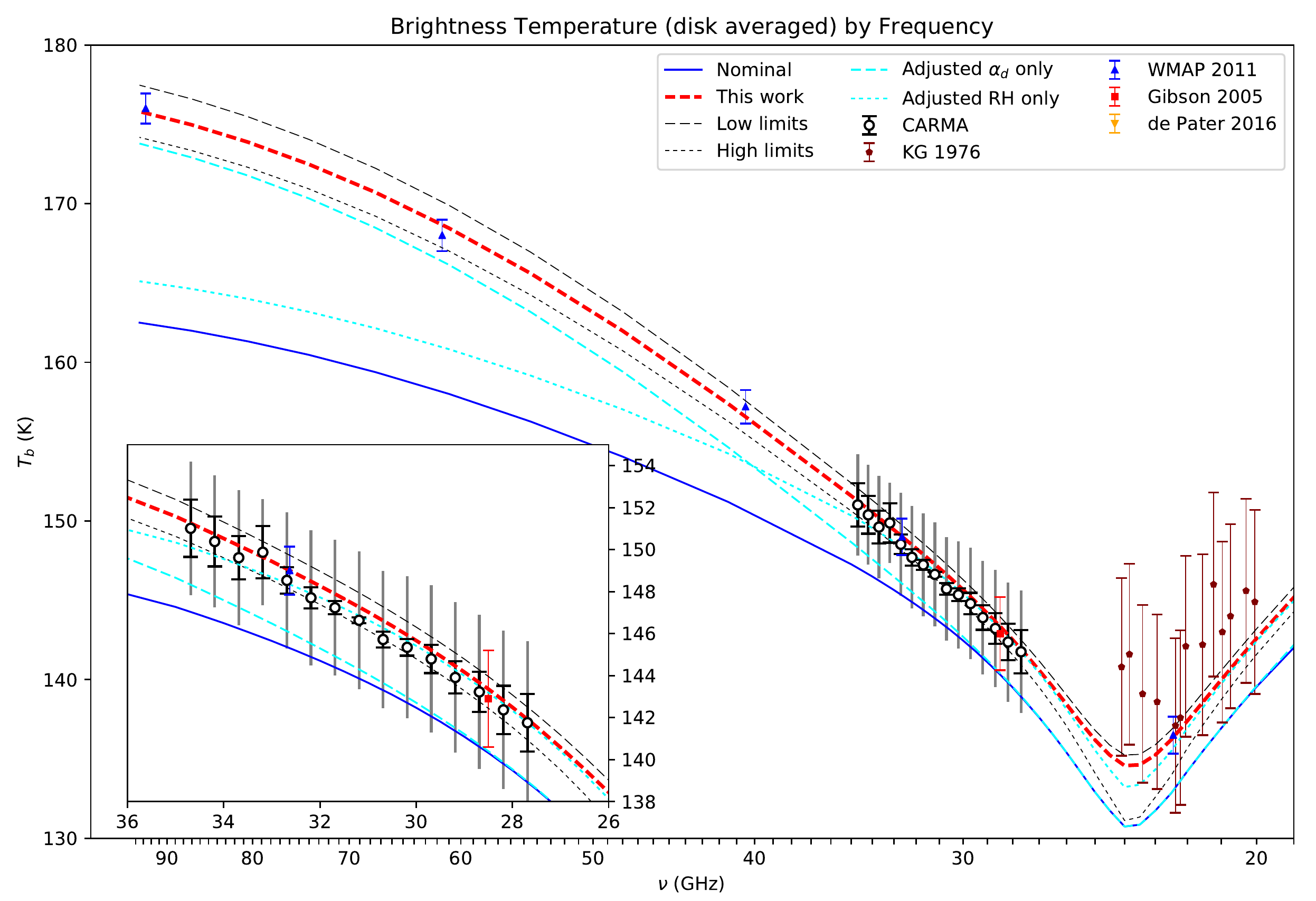}
	\caption{\label{fig:emission_freq}Similar to Figure \ref{fig:emission_freq}, but restricted to a smaller spectral range and set against a frequency axis. Inset panel provides a clearer view of the CARMA points reduced in this work. We include two additional models, indicated by cyan lines, representing one-parameter deviations from the nominal model; in other words, we hold two parameters at 100\% of the nominal value and let the third take on the preferred value based on the $\chi^{2}$ fit so as to demonstrate the effects of the parameters as well as their general independence to each other. The effects of relative humidity and deep atmosphere abundance are shown by the two cyan lines. High altitude atmosphere abundance is not shown here, but when decreased raises the brightness temperature at the band center and accounts for the band center difference between the relative-humidity-only and preferred (red-dashed) models.}
\end{figure*}

Our final model is shown by the red dashed lines in Figures \ref{fig:emission_wl} and \ref{fig:emission_freq}, with upper and lower bounds.
In Figure \ref{fig:emission_freq}, as mentioned above, we also show the spectra resulting from $\alpha_{d}$ and $RH$ only (cyan dashed and dotted lines, respectively).

\section{Conclusion}

We utilized data from the COPSS survey (\citealt{2015ApJ...814..140K}; \citealt{2016ApJ...830...34K}) at frequencies between 27-35 GHz to identify and extract observations of Jupiter, used as a secondary calibrator over the course of 5 months. These data were reduced into 15 frequency measurements of Jupiter's disk-averaged thermal brightness temperature over a relatively unobserved $\sim$10 GHz wide section of Jupiter's thermal emission profile just short of the NH$_{3}$ absorption band.
CARMA's strong spectral stability, and hence the observed slope in the spectrum, as well as the high precision in our absolute calibration, were key in deriving the disk-averaged NH$_{3}$ profile by fitting the data using our radiative transfer models.
We find that the NH$_{3}$ abundance below the NH$_{3}$ ice cloud, at $P \sim 8$ bar,  is $2.40\e{-4}$, bounded between $[2.26, 2.57] \e{-4}$, and carries up through the cloud reductions to $1.89\e{-4}$ at $P \sim 0.8$ bar.
Relative humidity, within the NH$_{3}$ cloud layer where the abundance follows the saturation curve, is found to be $56.5\%$, bounded between $[50\%, 63.5\%]$.
At high altitudes, well above the NH$_{3}$ cloud layer, the NH$_{3}$  abundance is near $2\e{-8}$, with  an upper bound of $2.4\e{-7}$.
These results echo the conclusion made in JG, especially that of sub-saturation by a factor of 2, and by \cite{2001Icar..149...66D, 2016Sci...352.1198D}.

This measurement set will be useful in future explorations, as have WMAP, JG, and others been in ours. They will, in a sense, extend the Juno data, since they are at shorter wavelengths than the MWR experiment on Juno.

CARMA's calibration is strong enough that these reduced thermal measurements may be useful as calibration information for interferometers lacking short baselines.
The measurements produced in this work are believed to encompass all flux from the disk, so they may be treated as ``single-dish'' flux measurements to give calibration context to interferometer scans.

\acknowledgements
We gratefully acknowledge the James S. McDonnell Foundation, the National Science Foundation (NSF) and the University of Chicago for funding to construct the SZA. Partial support was provided by NSF Physics Frontier Center grant PHY-0114422 to the Kavli Institute of Cosmological Physics at the University of Chicago. Support for CARMA construction was derived from the states of California, Illinois, and Maryland, the James S. McDonnell Foundation, the Gordon and Betty Moore Foundation, the Kenneth T. and Eileen L. Norris Foundation, the University of Chicago, the Associates of the California Institute of Technology, and the National Science Foundation (NSF). Support for CARMA operations and analysis was provided in part by the National Science Foundation University Radio Observatories Program, including awards AST-1140019 (to the University of Chicago), AST-1140031 (University of California-Berkeley), AST-1140021 (California Institute of Technology), and by the CARMA partner universities. In addition, partial support to this particular research endeavour was provided by NASA Planetary Astronomy (PAST) Award NNX14AJ43G to the University of California, Berkeley.

\software{pyplanet \citep{2014Icar..237..211D}}

\bibliographystyle{aasjournal}
\bibliography{sza_jupiter}

\end{document}